\documentstyle[12pt]{article}
\newcommand{\xq}{\begin{equation}}
\newcommand{\zq}{\end{equation}}
\newcommand{\beq}{\begin{eqnarray}}
\newcommand{\eeq}{\end{eqnarray}}
\begin{document}

\begin{center}
\hspace{50mm}
Preprint -94-23,\\
\hspace{60mm}{\sl Sov. Phys. JETP} {\bf 107} No.6 (1995)
\vspace{5mm}\\
{\large\bf On the superconductivity of 2D system
with arbitrary carrier density in external magnetic field}
\vspace{5mm}\\
{\sl V.P. Gusynin, V.M.Loktev  and I.A. Shovkovy}
\vspace{5mm}\\
{\sl Bogolyubov Institute for Theoretical Physics,\\
252143 Kiev, Ukraine}
\end{center}

\section*{Abstract}

Selfconsistent equations which describe the order parameter and
chemical potential behaviour in $2D$ metalic system as functions of external
magnetic field, $H$, temperature, $T$, and carrier density, $n$, are
obtained. It is shown that for the case of the local pairs (low
carrier density and negative chemical potential $\mu$) the derivative
$dH_{c_2}(T)/dT$ at $T=T_c$ is essentially less then that for the system
with Cooper pairs (high carrier density and positive chemical potential
$\mu\gg T_c$).
It is found that in magnetic field satisfying the quantum limit criterium
the system is characterized by non-trivial inhomogeneuos order parameter
which can exist at rather high temperatures.

\section{Introduction}

Despite of the great deal of efforts directed toward solving the problem,
the nature of high temperature
superconductivity (HTSC) retains to be unknown. However, those efforts have
not been done in vain. At present one can observe a common consent about
the most characteristic properties of HTSC \cite{Gor,Anders}.
Such properties are: {\em a)}
quasi--two--dimensional behaviour of the conductivity in the normal phase;
{\em b)} relatively low (at least considerably lower than in ordinary
metals) density of carriers. There are also many other distinctive
properties which may appear to be very important for understanding the
nature of HTSC. But it is unlikely to describe all the characteristic
features of HTSC in the framework of the only theoretical approach. So,
very often the role of one of them is studied and then the total picture
is created by means of summing up all the results in some "artistic" way.

One of the very important and intensively studied questions in the
theory of superconductivity (SC) is the question about a dependence of SC
properties of a system on the carrier density. The first who emphasized
on the relevance of such a problem was Legget \cite{Leg}. Later this
problem was also discussed in Refs.\cite{Eagles,Ran2,Gorbar}. In
Ref.\cite{Ran1} the existence of a crossover driven by the carrier density
was clearly established. In that paper the authors have considered a three
dimensional model with a local attractive interaction and shown that the
gradual transition from low to high carrier density is accompanied by a
crossover from local pair SC to the SC with Cooper pairs. As for HTSC, an
intermidiate case seems to be realized in it.

In this paper we continue the analysis of the papers \cite{Gorbar} and
\cite{Gorbar1}.

\section{Model and general discussion}
As in the papers \cite{Gorbar,Gorbar1}, we shall focus our attention only
on the qualitative side of the problem. So, for our purposes, we can
take the simplest Hamiltonian which describes a system of charge carriers
(fermions) with a local attraction between them:
\beq
\hat{H}&=&\int d^2 {\bf r} {\hat{\cal H}}({\bf r}),\\
{\hat{\cal H}}({\bf r})&=&-\psi^{\dagger}_{\sigma}
\left[\frac{1}{2m}\left(\partial_{j}-i\frac{e}{c}A_{j}\right)^2+\mu\right]
\psi_{\sigma}-\nonumber\\
&-&\frac{g}{2}(1-\delta_{\sigma\sigma_1})
\psi^{\dagger}_{\sigma}\psi^{\dagger}_{\sigma_1}\psi_{\sigma_1}\psi_{\sigma},
\label{eq:ham1}
\eeq
where $A_{j}$ is a vector potential corresponding to a magnetic field, $H$,
(note that in two dimensions magnetic field is pseudoscalar function), $e$
and $m$ are charge and mass of the carriers, respectively, $g>0$ is an
attractive coupling between these carriers and $\mu$ is the chemical
potential.

Introducing Nambu notations for fermion field
$\Psi=(\psi^{\dagger}_{\uparrow},\psi_{\downarrow})$ \cite{Nambu}, we can
rewrite (\ref{eq:ham1}) in a more convenient form:
\beq
{\hat{\cal H}}({\bf r})&=&-\Psi^{\dagger}\tau_{3}
\left(\frac{{\cal D}^2}{2m}+\mu\right)\Psi
+g\Psi^{\dagger}\tau_{+}\Psi\Psi^{\dagger}\tau_{-}\Psi,
\label{eq:ham2}
\eeq
where $\tau_3$, $\tau_{+}\equiv (\tau_1+i\tau_2)/2$,
$\tau_{-}\equiv (\tau_1-i\tau_2)/2$ are Pauli matrices and the covariant
derivative has the form ${\cal D}_{j}\equiv\partial_{j}-ie\tau_{3}A_{j}/c$.
Then the partition function is expressed through the Hamiltonian as:
\xq
Z\equiv \exp(-\Omega/T)=Tr\exp(-\hat{H}/T),\label{eq:partf1}
\zq
Since the thermodynamical potential, $\Omega=\Omega(V,T,\mu)$, is the
function of the chemical potential (besides the volume and the temperature)
and we are interested in a dependence of all physical values on the density
of the carriers, it is necessary to write down the second equation which
links the density, $n$, and the chemical potential, $\mu$:
\xq
n=\frac{1}{V}\frac{\partial \Omega}{\partial \mu}.\label{eq:dens1}
\zq
The expression for the partition function  with the Hamiltonian
(\ref{eq:ham2}) can be represented in a path integral form:
\beq
Z=\int[d\Psi^{\dagger}d\Psi]\exp\bigg[-\int\limits_{0}^{\beta}d\tau\int
d^2{\bf r}\left(\Psi^{\dagger}\partial_{\tau}\Psi+{\hat{\cal H}}({\bf r})
\right)\bigg],
\qquad \beta\equiv \frac{1}{T},
\eeq
where Grasmann variables $\Psi(\tau;{\bf r})$ satisfy the antiperiodic
boundary condition $\Psi(0;{\bf r})=-\Psi(\beta;{\bf r})$. After usual
introduction of an auxiliary scalar field by means of the
Habbard--Stratanovich trick:
\beq
Z&=&\int[d\Psi^{\dagger}d\Psi d\Phi d\Phi^{\ast}]\exp\bigg[
-\int\limits_{0}^{\beta}d\tau\int d^2{\bf r}
\bigg(\frac{|\Phi|^2}{g}+\nonumber\\
&+&\Psi^{\dagger}\left[\partial_{\tau}-\tau_3
\left(\frac{{\cal D}^2}{2m}+\mu\right)+\tau_{-}\Phi+\tau_{+}\Phi^{\ast}
\right]\Psi\bigg)\bigg],
\eeq
we can perform (at least formally) the integration over Grasmann variables and
represent the result through the effective action:
\xq
S_{eff}(\Phi,\Phi^{\ast})=-TrLnG^{-1}+\frac{1}{g}\int\limits_{0}^{\beta}
\int d^2{\bf r}|\Phi|^2,
\zq
depending only on the
scalar field $\Phi$. However, the next integration over scalar field we
can perform only approximately, for example, using so called saddle point
formalism. In our problem the "saddle point" is defined by the equation:
\xq
\frac{\delta S_{eff}(\Phi,\Phi^{\ast})}{\delta \Phi^{\ast}(\tau;{\bf r})}
=tr[G(\tau,\tau;{\bf r},{\bf r})\tau_{+}]+\frac{\Phi}{g}=0,\label{eq:crline}
\zq
where Green's function, $G$, of interacting fermions is defined as the
solution of the equation:
\xq
\bigg[-\partial_{\tau_1}+\tau_3\left(\frac{{\cal D}^2}{2m}+\mu\right)
-\tau_{-}\Phi-\tau_{+}\Phi^{\ast}\bigg]G(\tau_1,\tau_2;{\bf r_1},{\bf r_2})=
\delta(\tau_1-\tau_2)\delta({\bf r_1}-{\bf r_2}),\label{eq:gre}
\zq
with the boundary condition $G(\tau_1+\beta,\tau_2;{\bf r_1},{\bf r_2})
=-G(\tau_1,\tau_2;{\bf r_1},{\bf r_2})$.

Let $\bar{\Phi}$ be a solution to the equation (\ref{eq:crline}), then
the partition function in the next to the leading approximation takes a
form of the Gauss type path  integral:
\beq
Z&=&\exp(-\bar{S}_{eff})\int[d\tilde{\Phi} d\tilde{\Phi}^{\ast}]\exp\bigg[
-\int\limits_{0}^{\beta}d\tau_1\int\limits_{0}^{\beta}d\tau_2\nonumber\\
&\cdot&\int d^2{\bf r}_1\int d^2{\bf r}_2
\tilde{\Phi}^{\ast}(\tau_1;{\bf r}_1)
\Gamma^{-1}(\tau_1,\tau_2;{\bf r}_1,{\bf r}_2)\tilde{\Phi}(\tau_2;{\bf r}_2)
\bigg],
\eeq
where $\bar{S}_{eff}\equiv S_{eff}(\bar{\Phi},\bar{\Phi}^{\ast})$. New
field $\tilde{\Phi}$ describes fluctuations and $\Gamma$ is its propagator:
\beq
\Gamma^{-1}(\tau_1,\tau_2;{\bf r}_1,{\bf r}_2)&=&\frac{1}{g}
\delta(\tau_1-\tau_2)\delta({\bf r_1}-{\bf r_2})+\nonumber\\
&+&tr\left.\bigg[G(\tau_1,\tau_2;{\bf r}_1,{\bf r}_2)\tau_{-}
G(\tau_2,\tau_1;{\bf r}_2,{\bf r}_1)\tau_{+}\bigg]\right|_{\bar{\Phi}}.
\label{eq:gam}
\eeq
For the partition function in this approximation we have
\xq
Z=\exp(-\bar{S}_{eff}-TrLn\Gamma^{-1}).
\zq
So, as follows from Eqs.(\ref{eq:partf1}) and (\ref{eq:dens1}), the
expression for the carrier density takes the form of a sum of two different
terms:
\xq
n=\frac{T}{V}Tr[\tau_3G]-\frac{T}{V}\frac{\partial}{\partial\mu}
(TrLn\Gamma^{-1}).     \label{eq:dens2}
\zq
The first term is expressed through the fermion propagator and the second
through the propagator of scalar fluctuations. Having such a representation,
we shall refer to the first term as fermion part in density of the carriers
and to the second as boson one. The fact that the
ratio of fermions and composite bosons (fluctuations can also be interpreted
as a field of two--fermion composite particles) is  determined by the
dynamics is the main virtue of this model.

At this point we have, in principle, a closed selfconsistent system of
two equations which completely describe a dependence of the order parameter
and the chemical potential as functions of the temperature and carrier
density.

\section{On the critical line equation}
As was indicated in the previous section Eqs.(\ref{eq:crline}) and
(\ref{eq:dens2}) completely describe the behaviour of the order parameter
and the chemical potential of the system as a function of "external"
parameters $T$, $B$ and $n$. But analizing these equations in general case
is an unsolvable problem. So we restrict ourselves only to analizing the
behaviour of the system near the critical line (related to the second
type phase transition). Such a choice simplifies the problem considerablly
since when the system is in the nearcritical region we have a natural
small value (order parameter) and as a consequence  we can apply
pertubation theory in this value.

In order to simplify the problem further we assume that the solution to
the equation (\ref{eq:crline}) does not depend on "time" coordinate
$\tau$. In this case, as follows from Eq.(\ref{eq:gre}), Green's function
depends only on the difference of "time" variables $(\tau_1-\tau_2)$. So,
after taking into account the boundary conditions, Green's function can
be expanded into the Fourier series:
\xq
G(\tau_1-\tau_2;{\bf r}_1,{\bf r}_2)=T\sum_{n=-\infty}^{\infty}
G_{n}({\bf r}_1,{\bf r}_2)\exp[-i\omega_{n}(\tau_1-\tau_2)],
\qquad \omega_{n}=\pi T(2n+1).
\zq
Solving the equation for the Green's function (\ref{eq:gre}) in linear
approximation in order parameter and substituting it into (\ref{eq:crline}),
we come to the following integral equation:
\beq
\bar{\Phi}({\bf r})&=&\int d^2{\bf r_1}K_1({\bf r},{\bf r}_1)
\bar{\Phi}({\bf r}_1),\\
K_1({\bf r},{\bf r}_1)&=&-gT\sum_{n=-\infty}^{\infty}tr\bigg[
G^{(0)}_{n}({\bf r},{\bf r}_1)\tau_{-}G^{(0)}_{n}({\bf r}_1,{\bf r})
\tau_{+}\bigg],\label{eq:linear}
\eeq
where $G^{(0)}_{n}({\bf r},{\bf r}_1)$ denotes the Green's function at
$\Phi=0$ which can be easily found using the Schwinger proper time method
\cite{Sch}. Here we write down this function without deriving (for
details see appendix A in Ref.\cite{GLSh}):
\beq
G^{(0)}_{n}({\bf r}_1,{\bf r}_2)&=&\exp\left(-\frac{i}{2l^2}
\tau_3(x_1y_2-y_1x_2)\right)G^{(hom)}_{n}({\bf r}_1-{\bf r}_2),\\
G^{(hom)}_{n}({\bf r})&=&\frac{1}{2\pi l^2}\exp
\left(-\frac{{\bf r}^2}{4l^2}\right)
\sum_{j=0}^{\infty}L_{j}\left(\frac{{\bf r}^2}{2l^2}\right)\frac{1}
{i\omega_{n}-\tau_{3}[\omega_{H}(j+1/2)-\mu]}.\label{eq:green0}
\eeq
Strictly speaking, Eq.(\ref{eq:linear}) which has the form of a spectral
problem is valid only on the critical line where the order parameter
equals zero. The condition indicating the existence of a solution to this
equation determines the critical line in the problem. If we were going to
leave the critical line, we would need to take into account the next
nonlinear term. In this paper we are interested only in the behaviour on
the critical line.

After substituting the explicit expression for the Green's function into
(\ref{eq:linear}) we get:
\beq
\bar{\Phi}({\bf r})&=&\int d^2{\bf r_1}K^{(hom)}_1({\bf r}-{\bf r}_1)
\exp\left(\frac{i}{l^2}(xy_1-yx_1)\right)\bar{\Phi}({\bf r}_1),\label{eq:cr}
\eeq
where $l$ is the magnetic length and the homogeneous kernel, $K^{(hom)}_1$,
determined by
\beq
K^{(hom)}_1({\bf r},{\bf r}_1)&=&-gT\sum_{n=-\infty}^{\infty}tr\bigg[
G^{(hom)}_{n}({\bf r})\tau_{-}G^{(hom)}_{n}({\bf r})
\tau_{+}\bigg].
\eeq
As was shown in \cite{Rajag,Tesan}, Eq.(\ref{eq:cr}) has the exact
solution:
\xq
\bar{\Phi}({\bf r})=\Delta\exp\left(-\frac{{\bf r}^2}{2l^2}\right),
\label{eq:sol}
\zq
where in the simplest case $\Delta$ is constant. After substituting
(\ref{eq:sol}) into (\ref{eq:cr}) we come to the condition determining
the critical line:
\beq
1&+&\frac{gTl^2}{(2\pi)^3}\sum_{n=-\infty}^{\infty}\int\int
d^2{\bf K}d^2{\bf k}\exp\left(-\frac{{\bf K}^2l^2}{2}\right)\nonumber\\
&\cdot& tr\bigg[G^{(hom)}_{n}\left(\frac{{\bf K}}{2}+{\bf k}\right)\tau_{-}
G^{(hom)}_{n}\left(-\frac{{\bf K}}{2}+{\bf k}\right)\tau_{+}\bigg]=0.
\label{eq:crl}
\eeq

The second equation which links the density and the chemical potential is
the following:
\xq
n=\frac{T}{(2\pi)^2}\sum_{n=-\infty}^{\infty}\int d^2{\bf k}
tr[\tau_3G^{(hom)}({\bf k})]-\frac{T}{V}\frac{\partial}{\partial\mu}
(TrLn\Gamma^{-1}_{0}).     \label{eq:dens3}
\zq
where $\Gamma_{0}$ is defined by the expression similar to (\ref{eq:gam}) but
with Green's functions at $\Phi=0$.

Eqs.(\ref{eq:crl}) and (\ref{eq:dens3}) selfconsistently describe the
critical parameters $\mu(T)$ and $H_{c_2}(T)$ at every given carrier
density. Below we shall consider two limiting cases of low
and high magnetic fields.

\section{Low magnetic field}

This case is realized when the cyclotron frequency is much lower then
the temperature, $\omega_{H}\ll T_{c}$. Expanding Eqs.(\ref{eq:crl}) and
(\ref{eq:green0}) in power seriers in $l^{-1}$ and keeping only the lowest
term in magnetic field, we get the equation:
\xq
\frac{4\pi}{gm}=\int\limits_{0}^{W}\frac{du}{u-\mu}\tanh\frac{u-\mu}{2T}
-\frac{4T}{ml^2}\sum_{n=0}^{\infty}\int\limits_{-\mu}^{W-\mu}\frac{du}{u}
\frac{d}{du}\frac{u^2(u+\mu)}{\omega_{n}^{2}+u^2}, \label{eq:crfield}
\zq
where $W$ is the band width.

The critical temperature, $T_{c}$, when the field is swiched off is defined
by the equations:
\xq
\frac{4\pi}{gm}=\int\limits_{0}^{W}\frac{du}{u-\mu_c}\tanh
\frac{u-\mu_c}{2T_c}, \qquad \mu_c=\mu(T_c), \label{eq:crtem}
\zq
which have been analyzed in detail in Ref.\cite{Gorbar1}. Substituting
the expression for the coupling (\ref{eq:crtem}) into (\ref{eq:crfield})
and taking the limit $W\to \infty$, we come to:
\xq
\int\limits_{0}^{\infty}du\left[\frac{\tanh(u-\mu)/2T}{u-\mu}-
\frac{\tanh(u-\mu_c)/2T_c}{u-\mu_c}\right]=\frac{2T}{ml^2}
\sum_{n=0}^{\infty}\left[\frac{1}{\omega_{n}^{2}}+
\frac{\pi\mu}{2\omega_{n}^{3}}+
\frac{\mu\arctan(\mu/\omega_{n})}{\omega_{n}^{3}}\right], \label{eq:ttc}
\zq
Approaching the critical line at low but nonzero magnetic field,
we should come to the value of the temperature close to $T_c$. So,
expanding the left hand side of the Eq.(\ref{eq:ttc}) in $(T-T_c)$
we can find the slope of the critical line at $T_c$:
\xq
\frac{e}{4mc}\left(\frac{dH_{c_2}}{dT}\right)_{T_c}=
\frac{1+\tanh(\mu_c/2T_c)-(\mu_c/T_c)(\partial\mu_c/\partial T_c)
\tanh(\mu_c/2T_c)}{1+[7\zeta(3)/2\pi^2](\mu_c/T_c)+(|\mu_c|/2T_c)
\int\limits_{0}^{|\mu_c|/2T_c}(du/u^3)(u-\tanh u)}. \label{eq:slope}
\zq
Even though the slope of the critical line is given by the
explicit expression (\ref{eq:slope}) this does not solve the problem
completely because, as was indicated above, we should add also the
second equation (\ref{eq:dens3}) linking the chemical potential with
the carrier density. From that equation we should also find the
derivative
$\partial\mu_c/\partial T_c\equiv (\partial\mu/\partial T)_{T_c}$ which
is calculated at constant $n$. Since the analysis of the second equation
for arbitrary $\mu$ is quite a difficult problem we shall consider only
limiting cases.

\underline{1.Local pairs}. At first we consider the case of low density
of carriers and strong interaction between fermions, {\em i.e.}
$\epsilon_F\ll |\epsilon_b|$ where $|\epsilon_b|=2W\exp(-4\pi/mg)$ is the
energy of bound state in two particle problem and $\epsilon_F\equiv n\pi/m$.
Analyzing the propagator
of the scalar field describing fluctuations we come to the conclusion that
in the lowest approximation in $l^{-1}$ the propagator should be taken
at zero magnetic field (for details see \cite{GLSh}). Assuming that
$\mu_c<0$ and $|\mu_c|\gg T_c$ (the equivalence of these conditions and
the inequality stated above can be easily established), it is not difficult
to find the expression for carrier density in this limiting case:
\xq
n=-\frac{2mT_c}{\pi}\ln\left[1-\exp\left(\frac{2\mu_c-\epsilon_b}{T_c}
\right)\right]+O\left(\exp\left(-\frac{|\mu_c|}{T_c}\right)\right),
\label{eq:local}
\zq
where we omitted exponentially small terms.

In principle, Eqs.(\ref{eq:slope}) and (\ref{eq:local}) completely
describe the critical line slope at $T_c$, if the critical temperature
at zero field is nonzero and the density of carriers is a finite constant.
However, it can be easily shown that equations (\ref{eq:crtem}) and
(\ref{eq:local}) are consistent only
at $T_c=0$. This fact is in full agreement with the general
statements about the role of fluctuations in low dimentional
models with short--range interaction where, as is known,  there is
no room for a nontrivial order parameter \cite{Mermin}.
This problematic situation
is not actual in real HTSC since they are not strictly two--dimensional,
they are only quasi--two--dimentional. So, we come to the place where
we are not able to use a strictly two--dimensional model. In order to
avoid this obstacle we assume that the model will be treated correctly
if we take into account quasi--two--dimensional character of the model
only into the equation for carrier density which is extremely sensitive
to the number of dimensions and leave the equation (\ref{eq:slope})
without changes.

Since in three dimensions the density is measured in different units,
our enlargement of the phase volume of the system should be accompanied by
a redefinition of the "two--dimentional" density in the form:
$n^{2D}=n^{3D}\bar{K}^{-1}_z$ where $n^{3D}$ is the density found in the
$3D$ anisitropic model and $\bar{K}_z$ is some characteristic momentum.
Chosing $\bar{K}_z^2/4M\simeq T$ where $M$ is an effective mass in
$z$--direction, we get:
\xq
n=\frac{2mT}{\pi^2}\int\limits_{0}^{\infty}du\frac{u^{1/2}}
{\exp[u+(\epsilon-2\mu)/T-1]}.
\zq
Now we find the slope of the critical line without problems:
\beq
-\frac{e}{4mc}\left(\frac{dH_{c_2}}{dT}\right)_{T_c}\simeq
\left(\frac{\mu_c}{T_c}\right)^2\exp\left(-\frac{|\mu_c|}{2T_c}\right),
\qquad |\mu_c|=\frac{|\epsilon_b|}{2}\gg T_c\sim\epsilon_F.
\eeq
Obtained expression shows that the slope of the critical line in the
case of composite bosons is very small what is in qualitative agreement
with the result of the paper \cite{Alexandrov}.

\underline{2.Cooper pairs}. When the density of carriers is high or the
attraction between fermions is relatively weak, {\em i.e.}
$\epsilon_F\gg \epsilon_b$, we come to the SC of Cooper pairs. In terms of
chemical potential this condition is equivalent to the inequality
$\mu_c\gg T_c$. Since the fluctuations are suppressed in any
three--dimensional model and the mean field approximation is reliable when
the density of carriers is high enough, we can assume that the same
statement
is correct in our "quasi--two--dimensional" model (though it cannot be
shown in a direct way). With such an assumption the bosonic term in
Eq.(\ref{eq:dens3}) can be omitted and we get:
\xq
n=\frac{mT}{\pi}\ln\left(1+\exp\frac{\mu_c}{T}\right)\sim\frac{\mu_c m}{T},
\zq
and for the slope of the critical line:
\xq
-\frac{e}{4mc}\left(\frac{dH_{c_2}}{dT}\right)_{T_c}\simeq
\frac{2\pi^2}{7\zeta(7)}\frac{T_c}{\mu_c}, \qquad\mu_c\simeq\epsilon_F\gg
T_c=\frac{\gamma}{\pi}\sqrt{2|\epsilon_b|\epsilon_F},
\zq
where $\gamma$ is the Euler constant.

So, the results received in this section show that the behaviour of the
critical line, in particular its slope at $T_c$ essentially depends on
the density of carriers in the system. At low density the slope is
small and with increasing the density the slope becomes bigger
(all other parameters are kept constant). Note that after achieving some
maximum value the slope turns to decreasing since in the regime of Cooper
pairs it decreases with increasing of the carrier density.

\section{Strong magnetic field}

When a magnetic field is so strong that it satisfies the condition,
$\omega_H\gg T_c$, we come to the so called quantum limit. It is obvious
that in a very strong magnetic field, only the lowest Landau level in
fermion spectrum plays an important role. So, omitting all those terms in
fermion Green's function which correspond to the higher Landau levels, we
get:
\xq
G_{n}^{(hom)}({\bf k})=\frac{2\exp(-{\bf k}^2l^2)}{i\omega_n
-\tau_3[\omega_H/2-\mu]}.          \label{eq:approx}
\zq
This, in its turn, means that the density of states at the lowest level
should be higher than the density of carriers in the system. It is also clear
that the higher density of carriers is in the system, the stronger magnetic
field should be applied in order to achieve the quamtum limit. The second
condition when the approximation of the lowest Landau level is reliable
is connected with the value of coupling constant. When the attraction between
fermions is very strong the substitution of the approximate expression
(\ref{eq:approx}) instead of an exact Green's function will not be
satisfactory since any strong interaction could considerably change
the spectrum.

At last, all analysis in this section is performed in the framework of
the mean field approximation. We beleive that the consideration
taking into account fluctuations in quantum limit would have been
worth--while problem, but as is easily seen from all our consideration,
at first one need to find some simple formalism for treating
quasi--two--dimensional models which is free from the obstacles incorporated
into all two--dimensional models. It could be more realistic to consider
an anisotropic 3D model, however as we can judge such a problem is much
more difficult than ours and could be solved only numerically.

Since the expression for Green's function is so simple, we come at once
to the equation describing the critical line:
\xq
T=\frac{\omega_H-2\epsilon_F}{2\ln(\omega_H/|\epsilon_b|)
\ln(\omega_H/\epsilon_F-1)},
\zq
where we assumed that $n<eH/\pi c$ (or $\epsilon_F<\omega_H$). The
obtained result is similar to the result of paper \cite{RasTes}.

\section{Conclusion}

In this paper we derived the system of selfconsistent equations describing
second type phase transition in the simplest "quasi--two--dimensional"
system with local attraction between fermions in an external magnetic field.
We have shown that when the density of carriers in the system is
relatively low the equation, linking the chemical potential and the carrier
density, is nontrivial ($\mu\neq \epsilon_F$) and, as a result, this is
reflected in the behaviour of the critical line (its slope at $T_c$
is small). In the opposite case of figh density of carriers the system
can be described in the framework of the mean field approximation (as in
BCS model).

Our results are in qualitative agreement with the results obtained in
\cite{Alexandrov} for composite Bose particles though analytical
behaviour is different.


The research was supported in part by the grant No.K5O100 of the
International Science Foundation and by the International Soros Science
Education Program (ISSEP) through the grant No.PSU052143.

\end{document}